\documentclass[12pt]{iopart}

\usepackage{epsfig}

\begin{document}

\title{Simulational study for the crossover in 
the generalized contact process with diffusion}

\author{W.G. Dantas and M.J. de Oliveira}

\address{Instituto de F\'{\i}sica,
Universidade de S\~{a}o Paulo,
Caixa Postal 66318
05315-970 S\~{a}o Paulo, S\~{a}o Paulo, Brazil}

\author{J.F. Stilck}
\address{Instituto de F\'{\i}sica, Universidade Federal Fluminense,
24210-340, Niter\'oi, RJ, Brazil}

\begin{abstract}
In a recent work, Dantas and Stilck studied a model that 
generalizes the contact process model with diffusion \cite{wgd1}. 
Our approach, based on the supercritical expansion, 
showed that for a weak diffusion regime the crossover exponent 
between the directed percolation and compact directed percolation 
universality classes was $\phi\approx 2$. However this 
approach did not work for reduced diffusion rates higher 
than $D\approx 0.3$, where $0\leq D\leq 1$ and $D=1$ corresponds to an
infinite diffusion rate. Thus, in 
the present work we estimate this crossover exponent 
for higher diffusion rates using a numerical simulation approach.
\end{abstract} 

\pacs{05.70.Ln, 02.50.Ga,64.60.Cn}

\maketitle

%-----------------------------------------------------------------------
\section{Introduction}
	
Recently, Takeuchi et. al, using an experiment based 
on turbulent liquid crystal, measured the critical 
exponents of the directed percolation (DP) universality class \cite{take}.
This work meant a conquest for an area whose research has been
a growing interest in the three last decades, but which still was
missing a clear experimental realization. Indeed, the study 
of systems with absorbing states tries to provide a scenario
still poor in the understanding about the phase transitions in the
nonequilibrium 
regime, where no general theory is available, in opposition to what
occurs in the equilibrium regime with the Gibbs ensembles theory.

Using particular models to investigate general patterns, 
one hopes to get an insight of a possible general theory 
for these nonequilibrium phase transitions.
The most studied model in this area is the contact process (CP), introduced
by Harris as a {\it {toy model}} for the spreading of an epidemics 
\cite{harris}. 
Generally defined in a lattice, the CP consists of sites that may be
occupied by particles (sick individuals) or empty
(healthy individuals). An empty site becomes occupied
with a transition rate proportional to the number of occupied nearest
neighbor sites, and
an occupied site may become empty with a unitary rate. 
The competition between the dynamical rules for creation 
and annihilation of particles generates a phase transition 
between a state called absorbing that is devoid of particles
and an active phase, where some particles may survive. The presence of an
absorbing state prevents detailed balance and, because of that, this model
is intrinsically a nonequilibrium model. An interesting feature of the CP model
is that even in the one-dimensional case a phase transition appears, 
this does not happen in equilibrium cases for models with short range
interactions. Besides, this phase 
transition is characterized by critical exponents that belong to 
the directed percolation (DP) universality class, and there is a
conjecture that all models  
with a phase transition between absorbing and active states, with a 
scalar order parameter, short range interactions and no conservation laws 
do belong to this class \cite{md99,h00}. 

To determine the universality class for a particular model is an endeavor
in this research area and this may be accomplished using numerical or
analytical tools that allow us to obtain, as precisely as possible,
the critical exponents. Although the most usual of this techniques 
are numerical simulations, semi-analytical approaches as series expansions
are successful in some cases, furnishing a good precision to
critical exponents values.
Supercritical series expansion, according the prescription by Dickman
and Jensen \cite{dj} 
were used by us in past works to study critical exponents, particularly
the crossover 
exponent between different universality classes in models that generalize
the CP. In a first work \cite{wgd}, we determined the crossover exponent
between the DP and Compact Directed Percolation (CDP) classes with a
good precision, 
using PDA approximants \cite{fk77} in order to analyze numerically the
series in two variables.
On the other hand, we also studied how this crossover exponent could
be affected 
by the presence of diffusion in the model \cite{wgd1}. 
However, this approach only works for values
of the diffusion rate up to $D=0.3$, where $0\leq D\leq 1$, $D=1$
being equivalent
to the infinite diffusion regime. The failure of the series expansion
method for higher diffusion rates probably
occurred because in order to obtain a definite series expansion we 
were forced to include the diffusion term in the part of the evolution
operator which is treated as a perturbation.

In the present paper we extend the results obtained
in a preliminary work \cite{wgd2}, where we intented to obtain 
the behavior of this crossover exponent for larger values 
of the diffusion rate using a numerical simulational
approach to determine the critical line as a function of this
rate. This numerical simulational approach follows the algorithm 
proposed by Grassberger and De La Torre \cite{gras} to explore the
time evolution 
of the system from an initial condition where a unique seed is present in the
system. Besides the crossover between CP and CDP classes, now we also
analyze the model in the neighborhood of the mean-field limit, 
but not very close to the previous multicritical point, determining
another crossover exponent, characterized by the behavior of the
critical line close to the infinite diffusion rate limit.

The paper is organized as follows: in the section \ref{mod}
we present the model, its characteristics and the previous results
obtained in the past works \cite{wgd1,wgd}. In the section \ref{sim} we discuss
the numerical simulation approach and
the results for the crossover exponents - between CP and CDP and CP and
the mean field regime. Finally, the conclusions may be found in
section \ref{conc}. 

\section{The model}
\label{mod}
In a one-dimensional lattice with periodic boundary condition
each site may be empty or occupied by a single particle,
with a occupation variable $\eta_i$, assuming values 0 or 1 if site
$i$ is empty or occupied,
respectively. Thus, a configuration of the system at time $t$
is expressed by the vector $|\eta\rangle=\otimes|\eta_i\rangle$. 
In a continuos time sense, the dynamical rules of this model are:

\begin{enumerate}
\item We choose randomly a site $i$.
\item If this site is empty, it becomes occupied with a
transition rate equal to $p_an_{occ}/2$, where $n_{occ}$ is the number
of particles in the first neighbor sites.
\item Otherwise, if this site is occupied, it may become empty by two
processes:
\begin{enumerate}
\item By a contact process with first neighbor empty sites with a
  transition rate 
$p_bn_{emp}/2$, where $n_{emp}$ is the number of holes in the
first neighbor sites.
\item Spontaneously, with a rate $p_c$.
\end{enumerate}
\item Also a diffusive process takes place with a transition rate 
$\tilde{D}$. 
\end{enumerate}

The parameters $p_a,p_b$ and $p_c$ are positive numbers and obey
the relation $p_a+p_b+p_c=1$ and $0\leq\tilde{D}<\infty$. For
convenience, we will discuss this model in the parameter space
$(p_a,p_c,\tilde{D})$. For $p_b=0$ this model is equivalent
to the CP with diffusion and if $p_c=0,p_a=p_b$, and $\tilde{D}=0$ it
corresponds to the linear Glauber model (or voter  
model)\cite{glaub}. In the
last case, the critical exponents belong to the CDP universality class
and when $p_c\ne 0$, DP exponents are expected. Therefore, we have a 
crossover between these two classes in the neighborhood of $p_c=0$
and a generic density $g$ in this region should have the
following scaling form

\begin{eqnarray}
\label{eq1}
g(\Delta p_a,\Delta p_c,\tilde{D})\sim
\Delta p_a^{e_g(\tilde{D})}F\left(\frac{\Delta p_c}{|\Delta p_a|^
{\phi(\tilde{D})}}\right),
\end{eqnarray}
where $\Delta p_a=p_a-1/2$, $\Delta p_c=p_c$, $e_g(\tilde{D})$ is
a critical exponent related with the density $g$ and $\phi(\tilde{D})$ is
the crossover exponent. Besides, the scaling function $F(z)$
is singular at a value $z_0(\tilde{D})$ of its argument, which
corresponds to the critical line for a certain diffusion rate value, 
$\tilde{D}$. Using the scaling (\ref{eq1}), we have that the critical
line is asymptotically given by 
$p_c=z_0(\tilde{D})\Delta p_a^{\phi(\tilde{D})}$. 

On the other hand, for a nonzero value of $p_c$ and with diffusion rate
$\tilde{D}\rightarrow\infty$, the critical behavior of the model tends
to the one predicted by the mean-field approximation. An asymptotic form
to the critical line for the region of very high diffusion rate would
therefore be given by the scaling relation

\begin{eqnarray}
(p_a-p_a^c)=f_0(D-D_c)^{\phi_{MF-DP}},
\end{eqnarray}
where $\phi_{MF-DP}$ is the crossover exponent between the DP
and mean-field critical behavior and $D=\tilde{D}/(1+\tilde{D})$. This
relation is valid 
for constant $p_b$. When $p_b=0$ we have a known result
obtained by Konno \cite{kon} and verified by us and Messer and 
Hinrichsen \cite{wgd4,messer}, with $\phi_{MF-DP}=3$.

Using a mean-field cluster approximation (three-site level), 
we have shown that the crossover exponent has the value $\phi=2$ in the
weak diffusion regime \cite{wgd1}, while a slight deviation 
to smaller values appears in an intermediate regime, converging to $\phi=1$
in the strong diffusion limit. On the other hand, the crossover
exponent $\phi_{MF-DP}$ 
value is unitary, being determined by two-site level approximation
which furnishes the following relation for the critical line

\begin{eqnarray}
(1-D_{eff})=\frac{(1-\alpha)(1+\alpha)}{1-\alpha[1-\xi\alpha+\alpha^2(1-\xi)]},
\end{eqnarray}
where $\alpha=(1-p_a)/p_a,
D_{eff}=\alpha\tilde{D}/(1+\alpha\tilde{D})$ and $\xi=1-p_b$. 

The results obtained using series expansions \cite{wgd1} have shown that, at
least up to $D=0.3$, 
where $D=\tilde{D}/(1+\tilde{D})$, the crossover exponent is
$\phi=2$. Unfortunately, 
we were not able to study larger values of the diffusion rate probably
because, to derive a well
defined series, the diffusive term of the temporal operator must be treated
as a perturbation, limiting the values of the diffusive transition rate that
can be analyzed with reasonable precision.

\section{Numerical simulations}
\label{sim}
\begin{figure}
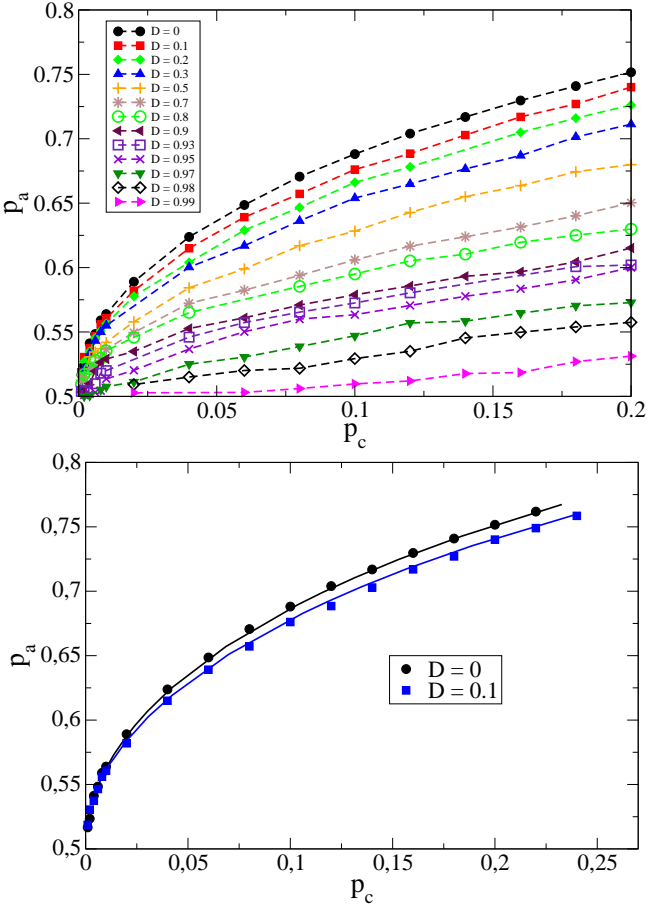

\begin{center}
\epsfig{file=./digfaseabdif.eps,scale=0.35}
\epsfig{file=./compsersimu.eps,scale=0.35}
\caption{Top: Phase diagrams obtained
using numerical simulation approach. 
Bottom: Comparative picture between the numerical simulation
and series expansion results to the same values of the diffusion 
rate.} 
\label{fig1}
\end{center}
\end{figure}

To simulate this model, we follow the scheme introduced by Grassberger and
De La Torre \cite{gras} to determine the critical point in models with
absorbing 
states through a numerical simulation of the time evolution. We carry it out
using a single seed particle as initial condition
and a lattice with $N=10000$ sites with periodic boundary
conditions. The time is discretized and at each step we realize the following
dynamical rules:
\begin{enumerate}
\item A list of all sites occupied by particles is stored, 
  and at each time step one of them is chosen randomly.
\item Once the site is chosen, a random number p uniformly 
  distributed in the interval [0, 1] is generated. 
  If $p<D=\tilde{D}/(1+\tilde{D})$ the particle jumps to one of the
  empty first neighbor sites,   
  if possible If both are empty, the destination site is chosen with
  equal probability. Otherwise, we choose a reaction: creation or annihilation.
  With a probability $p_a$ an empty site, in one of the first 
  neighbors, is occupied, if possible. Otherwise, the particle at 
  the chosen site will be annihilated either through 
  the spontaneous or through the auto-catalytic process.
\item To define the process switching a particle to a hole, 
  another random number $q$ is generated. If $q<pc/(1-pa)$ 
  the change is spontaneous, otherwise it will happen with 
  a probability proportional to the number of empty sites 
  in first neighbors of the chosen site.
\item The time interval associated with the steps above is 
  $\delta t = 1/N_A$, where $N_A$ is the number of sites occupied by 
  particles before the step. The process is repeated until 
  either a maximum number of steps $n_{max}$ is attained or the
  absorbing state $N_A=0$ is reached.
\item A total of $N_{rep}$ runs are done and mean values are calculated as
  functions of time. 
\end{enumerate}
In our simulations, we choose $n_{max}=2\times 10^5$ and $N_{rep}=2000$. Thus,
for a fixed value of $D$ and $p_c$, we estimate the the critical point
of the model, $p_a^c$, generating a sequence of critical curves as shown
in the figure \ref{fig1}. As we can see, there is a good
agreement between the simulational data and that one obtained by
series expansion. 
Not surprisingly, in the high diffusion limit, more fluctuations are
observed, preventing a precise determination for the critical point. 
This difficulty, in turn, is reflected in a poorer estimate of the
crossover exponent. 

\begin{figure}
\begin{center}
\epsfig{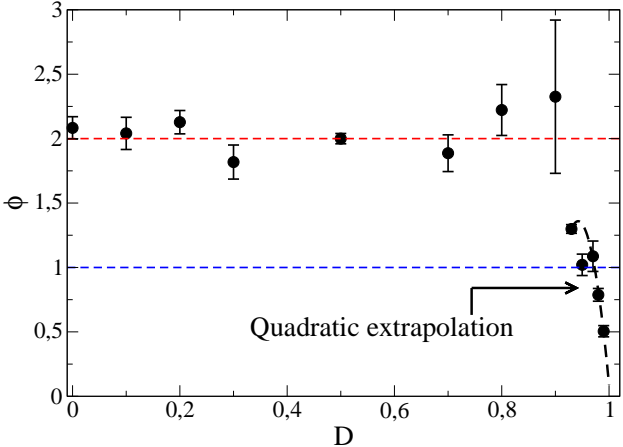}
\caption{Behavior of the crossover exponent as a function
of the diffusion rate.} 
\label{fig2}
\end{center}
\end{figure}

Using this curves for different diffusion rates
and the scaling form for a particular critical curve, $p_c=z_0\Delta
p_a^{\phi}$, 
we calculate how the crossover exponent behaves as a function of the
diffusion rate.
This is shown in the figure \ref{fig2}. The results fluctuate around
$\phi=2$ and, starting at $D=0.9$, we observe a systematic change of
the mean value of the estimates, 
converging approximately to $\phi=0$. Actually, an quadratic
extrapolation
results in $\phi=0.09$ when $D=1$, making reasonable the idea that we recover
the mean-field prediction of the critical line for infinite diffusion,
in accordance with the one-site level approximation result, where
$p_a=1/2, \forall p_c$.

\begin{figure}
\begin{center}
\epsfig{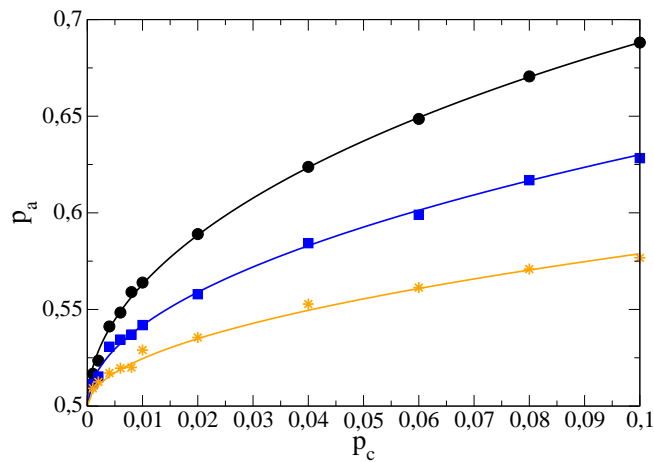}
\caption{Comparison between the simulational data (points) and
the critical lines obtained by expression $p_a=A(D)p_c^{1/2}+B(D)p_c^
{1/2+\varphi}$(full lines).} 
\label{fig3}
\end{center}
\end{figure}

These critical curves may be fitted with a good agreement by the
expression $p_a-1/2=A(D) p_c^{1/2}+B(D) p_c^{1/2+\varphi}$,
as shown in the figure \ref{fig3}, the 
values of the coefficients $A,B$ and the exponent $\varphi$ 
being given in the table \ref{tab1}. Notice that around $D=0.9$, $\varphi$
tends to zero and $|A|\rightarrow|B|$, but with opposite signals.
This behavior seems indicate the same tendency to $\phi=0$ exhibited
in the figure \ref{fig1}. However, we believe that the intermediate 
region where $0<\phi<2$ must be a numerical artifact, since in the
numerical simulational approach, 
a discontinuous transition in numerical values is not expected, so probably
we should have $\phi=2$, when $p_c\ne 0$ and $\phi=0$ otherwise.

\begin{table}[h!]
\caption{Coefficients of the expression which fits the critical lines.}
\label{tab1}
\begin{center}
\begin{tabular}{c c c c}  \hline\hline
D   &    A     &   B    & $\varphi$\\
\hline\
0   & 0.636    & -0.297 & 0.861\\
0.1 & 0.742    & -0.256 & 0.129\\
0.2 & 0.871    & -0.397 & 0.049\\
0.3 & 0.626    & -0.182 & 0.094\\
0.5 & 0.419    & -0.035 & 0.643\\
0.8 & 1.160    & -0.891 & 0.016\\
0.9 &-0.075    &  0.327 & 0.005\\
0.93&-3.478    &  3.738 & 0.004\\
0.95&-13.193   & 13.472 & 0.003\\
0.97&-56.760   & 56.983 & 0.001\\
0.99&-40.564   & 40.755 & 0.001\\
\hline\hline
\end{tabular}
\end{center}
\end{table}

\begin{figure}
\begin{center}
\epsfig{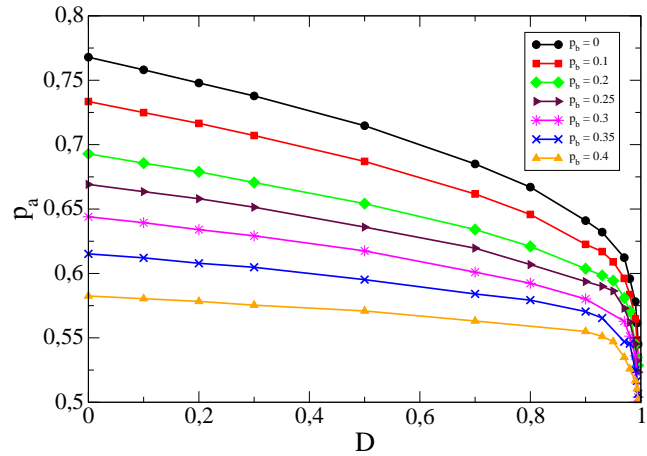}
\caption{Critical lines obtained for fixed values of$p_b$.} 
\label{fig4}
\end{center}
\end{figure}

\begin{figure}
\begin{center}
\epsfig{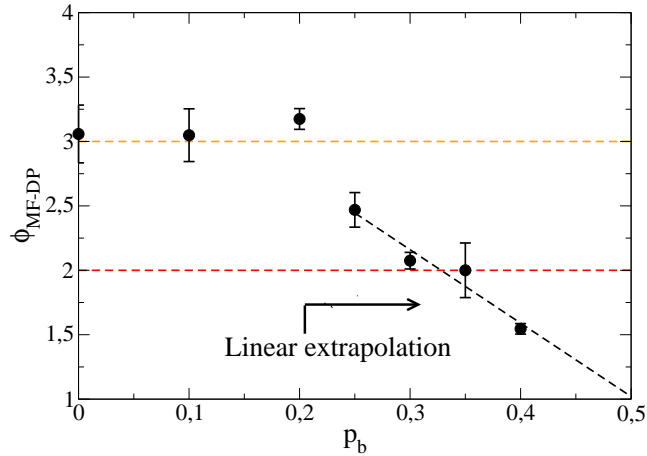}
\caption{Crossover exponent between the DP and mean-field regime.} 
\label{fig5}
\end{center}
\end{figure}

On the other hand, making $p_b$ constant, we can analyze how
the critical points behave in the neighborhood of the mean
field regime, which corresponds to an infinite diffusion rate limit, but not
so close of $p_c=0$. This
behavior for the contact process was considered exactly
by Konno \cite{kon} and more recently by us \cite{wgd4} and Messer and
Hinrichsen \cite{messer}, using series and field theoretic analysis,
respectively. 
For the CP the expected result is $\phi_{MF-DP}=3$, but to obtain
reliable estimates from simulations it
is needed to obtain critical points in a region very close
to the limit of infinite diffusion rate, since a secondary behavior 
emerges for high rate values before the asymptotic regime is reached,
as shown in \cite{fiore},  
where this exponent was estimated to be $\phi_{MF-DP}=4$. Our simulations
allowed us to reach values up to $D=0.995$ and the critical curves described
in the space $(D,p_a)$ are shown in the figure \ref{fig4}. Finally, 
analyzing the behavior of these curves close to the point $(D=1,p_a=1/2)$,
we found the crossover exponents $\phi_{MF-DP}$ exhibited in the
figure \ref{fig5}, where CP with diffusion is a particular case
that corresponds to $p_b=0$. We observe in a region near
to this point that the crossover exponent value is $\phi_{MF-DP}=3$.
However a trend to smaller values starts around $p_b=0.2$.
This tendency, in a linear extrapolation, seems result in an 
unitary exponent, already for $p_b=1/2$, according the prediction of the 
mean-field approximation. Actually, we obtain 
$\phi_{MF-DP}=1.02$ when $p_b=1/2$ with this extrapolation.

\section{Conclusions}
\label{conc}
In this paper we discuss the behavior of the two crossover exponents
for a model which generalizes the contact process with diffusion.
In a first moment we look for the behavior of the crossover exponent
that characterizes the change of DP to CDP universaltity classes. This
behavior is found in the neighborhood of the point $(p_a=1/2,p_c=0)$ for
a fixed diffusion transition rate. A preliminary result for this behavior
was accomplished using a perturbative series expansion method,
but estimates were restricted to values below $D=0.3$, with
$0\leq D < 1$ \cite{wgd1}. Here, using a numerical simulation 
of the time evolution,
we obtained a strong evidence that this exponent has the same value
of the case without diffusion, $\phi=2$, changing this value only in
the extreme case in which the diffusion rate diverges. In this
last case, we expected that the value of $\phi$ should vanish,
according to the 
mean-field approximation, where $p_a=1/2$, independently of $p_c$.
Besides, we speculate in a preliminary paper \cite{wgd4} that this
numerical change was related with a possible loss of the compact
clusters of particles in a high diffusion regime. Unfortunately this
possibility is still a speculation, since it is restricted
at this moment to a phenomenological argument. We also studied a
second model 
with a different dynamics for the diffusion, where the particles,
if possible, jump to a next nearest neighbor. Nevertheless, this version
furnished results very similar to an original case, so we
omit these results here.

On the other hand, the behavior of the model in a region not too near
$p_c=0$, with $p_b$ fixed, allowed us to estimate the crossover
exponent $\phi_{MF-DP}$, which characterizes the change of the critical
exponents from a DP to a mean-field critical regime, in the limit of
infinite diffusion limit. The original case for the CP with
diffusion, equivalent to set $p_b=0$, was already exactly studied by
Konno who obtained $\phi_{MF-DP}=3$. This value was re-obtained by
us and Messer and Hinrichsen using different approaches \cite{wgd4,messer}.
With the data originating of our simulations we concluded that
this exponent maintains the same value for $p_b\approx 0$, tending to
one
when $p_b\rightarrow 1/2$, in accordance with the mean-field approach
in its pair 
approximation. However, it should be remembered that when
$p_b=1/2$ the model belongs to 
CDP regime, and this could be a reason for this tendency. Finally, 
it would be interesting
to a more complete understanding of this model if these crossover
exponents would be 
related by a scaling relation, since the curves used to determine
these two exponents are not independent.

\section*{Acknowledgement}
W.G. Dantas acknowledges the financial 
support from  Funda\c{c}\~ao de Amparo \`a Pesquisa do
Estado de S\~ao Paulo (FAPESP) under Grant No. 05/04459-1 and
JFS acknowledges financial support by CNPq.

\end{document}